 \documentclass[final,5p,times,twocolumn,authoryear]{elsarticle}

\usepackage{amssymb}
\usepackage{lipsum}
\usepackage{url}
\usepackage{breakurl}
\usepackage{amsmath}
\usepackage{tikz}

\usepackage{upgreek}
\usepackage{hyperref, cleveref}
\usepackage[english]{babel}
\usepackage[utf8]{inputenc}
\usepackage{graphicx} 
\usepackage{caption}
\usepackage{subcaption}
\setcitestyle{square,comma,numbers}
\begin{document}

%\begin{frontmatter}



\title{Simulation-based performance comparison of varied pitch sizes GEM detectors}


\author[inst1]{Rajiv Gupta}

\affiliation[inst1]{organization={Department of Physics, Banaras Hindu University}, %Department and Organization
            %addressline={Varanasi}, 
            city={Varanasi},
            postcode={221005}, 
            state={U.P.},
            country={India}}

\author[inst1]{Gauri Devi}
\author[inst1]{Sunidhi Saxena}
\author[inst2]{Arpit Singh}
\author[inst1]{Ajay Kumar\texorpdfstring{\textsuperscript{*}}{}}
\cortext[*]{Corresponding author\\ Email: ajay.phy@bhu.ac.in}

\affiliation[inst2]{organization={Department of Physics, Indian Institute of Technology Bombay},%Department and Organization
            %addressline={Address Two}, 
            city={Mumbai},
            postcode={400076}, 
            state={Maharastra},
            country={India}}

\begin{abstract}
\noindent

Gas Electron Multiplier (GEM) detectors, typically featuring a standard pitch size of 140 $\mu$m and an inner hole diameter of 50 $\mu$m, are extensively utilized in high-energy physics experiments for tracking, triggering, and timing measurements. Their characteristics, such as high gain, good position resolution, improved temporal resolution, low discharge probability, radiation hardness, and high rate capabilities, make them highly favoured. Recent experimental studies have shown that triple-GEM detectors with a reduced pitch size of 90 $\mu$m and a smaller hole diameter of 40 $\mu$m can perform better than standard-pitch GEM detectors. To assess the effectiveness of these reduced dimensions, we conducted a simulation-based study using ANSYS and Garfield++. As a first step, we validated the simulation framework by modelling a standard single GEM detector and comparing the results with previous simulations and experimental data. Following validation, we designed GEM structures with reduced pitch sizes of 90 $\mu$m and 60 $\mu$m. We then performed a comparative analysis, focusing on key performance parameters like effective gain, electron transparency, and position resolution. These parameters were varied against an increase in GEM potential, drift electric field, induction electric field, drift gap, induction gap, and gas composition to optimize the performance of the detectors.

\end{abstract}
\maketitle

\section{Introduction}

\label{introduction}

Significant advancements in high-energy physics, such as dark matter searches, heavy-ion physics, neutrino physics, and the discovery of the Higgs boson, have been made possible by the development of particle accelerators and detectors. Such progress would have been difficult to pursue without the evolution of gaseous detector technologies. The year 1968 marked a revolution in gaseous detector technology when the Multi-Wire Proportional Counter (MWPC) was introduced by George Charpak for the first time \cite{Charpak:1968kd}. MWPC performed better than the traditional wire chambers and scintillating arrays. Charpak introduced an innovative approach of utilizing each wire between two parallel electrodes as a proportional counter. This idea remarkably enhanced the detection capabilities, which enabled the system to achieve good time resolution ($\sim$100 ns) and high counting rates ($\sim10^{5}$/wire). Owing to these advantageous characteristics over earlier bulky wire chambers and scintillators, MWPCs were quickly adopted in most particle physics experiments. Nevertheless, MWPCs have certain drawbacks despite their great qualities \cite{Sauli:2016eeu}. One of its drawbacks is the accumulation of positive ions at the cathode regions, which eventually lowers the detector's efficiency and gain. They are also ineffective in resolving multiple tracks in regions with high track density. The pollutants or organic buildup on the anode wire is another problem that can shorten the detector's life span and performance.

Anton Oed addressed some of these limitations in 1988 when he invented the Micro-Strip Gas Counter (MSGC) \cite{Oed:1988jh}. Instead of wire spacing for detection, MSGC uses microstrips fixed on a glass substrate for electron multiplication. This design improved rate capabilities to 2.3 MHz/cm², and the issue of multiple-track resolution was resolved. Despite such features, MSGCs are prone to early discharge, restricting their widespread applicability. This discharge problem occurs when the number of electron-ion pairs generated exceeds the
Raether limit ($\sim 10^{7}$ electron-ion pairs)~\cite{Fonte:1998vzw}.

Various efforts were made to address the discharge problem while preserving multi-track resolution and high-rate handling capabilities. Eventually, these efforts led to the development of advanced Micro Pattern Gas Detector (MPGD) designs \cite{CSEM_meeting}, such as the Micromesh Gaseous Structure (Micromegas) \cite{Charpak:2001tp}, the Micro Compteur à Trous (MicroCAT) \cite{Sarvestani:1999up}, and the Multi-Step Avalanche Chamber (MSAC) \cite{Charpak:1978tj}.

Micromegas comprise a double-stage parallel plate avalanche chamber with a narrow amplification gap (50–100 $\mu$m). This narrow gap causes an avalanche effect due to an intense electric field, which enables enhanced charge collection. It also possesses high-rate capability while achieving a gain on the order of $10^5$. On the other hand, MicroCAT has a structural design similar to Micromegas but consists of round holes instead of a micro-mesh. It was initially developed for 2D medical imaging of X-rays and exhibits important characteristics like high gain ($\sim 10^{4}$), good energy resolution, and high rate capability.

These detectors, like Micromegas and MicroCAT, work on the single-stage multiplication of electrons in high-intensity electric field regions, so they often suffer from discharge issues. Another disadvantage of using these detectors is the requirement for high-voltage operation. Hence, designing gaseous detectors with multi-stage multiplication could be a possible way to avoid such discharge issues. In such a configuration, each stage offers a relatively lower gain, but the overall setup can achieve sufficient gain from amplification. 

One such detector based on a multi-stage multiplication process is a Multi-Step Avalanche Chamber (MSAC) \cite{Charpak:1978tj}, which is widely used for single electron detection. In MSACs, an injected electron undergoes pre-amplification in the high electric field region between two meshes. The multiplied electrons from the avalanche effect drift towards successive cascaded high-field meshes, further amplifying and producing an enhanced signal at the readout electrode. This cascaded design enables higher gain ($\geq 10^{4}$) while preventing early discharge \cite{Sauli:2016eeu}. 

\begin{figure}
\begin{center}
\begin{tikzpicture}
    \filldraw[fill=pink!10, draw=black, thick] (-2,-3) rectangle (6,5);
    \coordinate (A) at (1,0);
    \coordinate (B) at (3,0);
    \coordinate (C) at (3.5,0.5);
    \coordinate (D) at (3,1);
    \coordinate (E) at (1,1);
    \coordinate (F) at (0.5,0.5);
    \node at (2,3) {\textbf{Drift Region (3 mm)}};
    \node at (2,-2) {\textbf{Induction Region (2 mm)}};
    \filldraw[fill=gray!40, draw=black, thick] (A) -- (B) -- (C) -- (D) -- (E) -- (F) -- cycle;
    \coordinate (A) at (-2,0);
    \coordinate (B) at (-1,0);
    \coordinate (C) at (-0.5,0.5);
    \coordinate (D) at (-1,1);
    \coordinate (E) at (-2,1);
    \filldraw[fill=gray!40, draw=black, thick] (A) -- (B) -- (C) -- (D) -- (E) --  cycle;
    \coordinate (A) at (5,0);
    \coordinate (B) at (6,0);
    \coordinate (C) at (6,1);
    \coordinate (D) at (5,1);
    \coordinate (E) at (4.5,0.5);
    \filldraw[fill=gray!40, draw=black, thick] (A) -- (B) -- (C) -- (D) -- (E) -- cycle;
    \filldraw[fill=red!80, draw=black, thick] (1,0) rectangle (3,-0.15);
    \filldraw[fill=red!80, draw=black, thick] (1,1) rectangle (3,1.15);
     \filldraw[fill=red!80, draw=black, thick] (-2,0) rectangle (-1,-0.15);
     \filldraw[fill=red!80, draw=black, thick] (-2,1) rectangle (-1,1.15);
     \filldraw[fill=red!80, draw=black, thick] (5,0) rectangle (6,-0.15);
     \filldraw[fill=red!80, draw=black, thick] (5,1) rectangle (6,1.15);
     \filldraw[fill=red!80, draw=black, thick] (-2,-3) rectangle (6,-3.15);
     \filldraw[fill=red!80, draw=black, thick] (6,5) rectangle (-2,5.15);
     \draw[<->, thick] (0,-1) -- (4,-1);
     \node at (2,0.5) {\textbf{\textsf{\scriptsize Kapton (50/50/25 \textmu m)}}};
     \node at (2,-0.9) {\textbf{Pitch=140/90/60 \textmu m}};
     \draw[<->, thick] (-0.5,0.5) -- (0.5,0.5);
     \draw[<-, thick] (0,0.5) -- (0,1.5);
\node at (0,1.5) {\textbf{50/40/25 \textmu m}};
\node at (2,1.3) {\textbf{\scriptsize{Metal}}};
\node at (2,2.25) {\textbf{Ar:$\textbf{CO}_{2}$(70:30)}};
\draw[<->, thick] (-1,-0.1) -- (1,-0.1);
\node at (0,0.11) {\textbf{\textsf{\scriptsize{70/55/30 \textmu m}}}};
\end{tikzpicture}  
\caption{Schematic representation of single GEM structures (SGEM, FGEM, FTGEM) illustrating the drift, hole, and induction regions. The diagram highlights the geometrical dimensions, including inner and outer hole diameters, Kapton layer thickness, and pitch for each configuration.}
\label{Schematic}
\end{center}  
\end{figure}
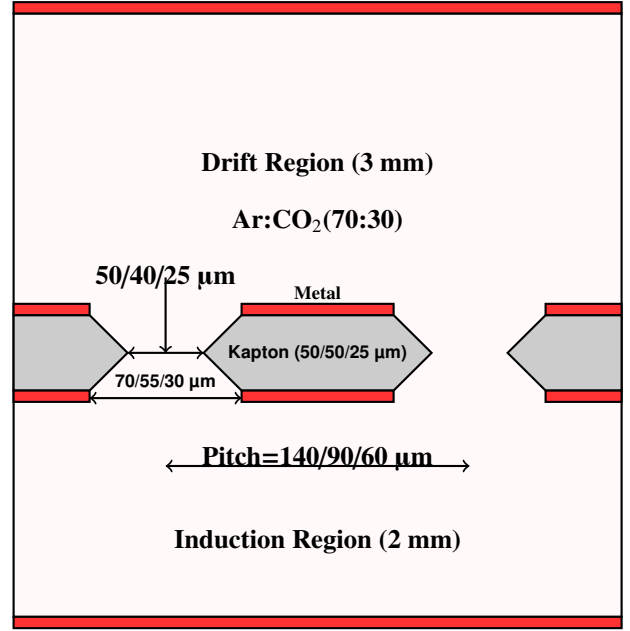

Inspired by the fundamental design of MSACs, F. Sauli developed the Gas Electron Multiplier (GEM) detector in 1997 \cite{Sauli:1997qp}. Figure \ref{Schematic} shows a schematic cross-section of a single GEM. The first GEM detector consisted of a 25 $\mu$m thick polymer Kapton layer sandwiched between 18 $\mu$m thick copper electrodes. They have a pitch size of 100 µm and outer hole diameters of 70 $\mu$m \cite{Sauli:1997qp}, which allowed for adequate charge amplification. A strong electric field is generated within the holes by applying a potential difference ($\Delta \text{V}_\text{GEM}$) across the copper electrodes. As the electrons accelerate toward these holes, they encounter an intense electric field that triggers an avalanche multiplication. As a result, some of the amplified electrons drift toward the subsequent GEM holes placed in a stacked configuration. Consequently, the enhanced signal is obtained at the readout electrode.

The amplification process in the GEM detector is influenced not only by its geometrical parameters but also by the composition and type of the gas mixture. The initial studies on GEM employed a gas mixture of Argon-dimethyl ether (Ar-DME) in a 90:10 ratio \cite{Sauli:1997qp}. However, subsequent experiments favoured uses of an Argon-Carbon dioxide \text{(Ar-CO\textsubscript{2})} mixture in a 70:30 ratio due to its non-flammable nature and cost-effectiveness \cite{Bachmann:1999xc}. This gas mixture is ideal for applications that require rapid signal processing due to the high drift velocity achieved by the electrons. Argon, with an ionization energy of 15.6 eV, is the primary ionizing gas in this mixture. It maintains a balance between charge multiplication and the stability of the detector. At the same time, CO\textsubscript{2} acts as a quencher that absorbs excess photons \cite{Mondal:2024ogs}.

Significant efforts have been made through experiments \cite{Barbeau:2004zr, Adak:2016avl} and simulations~\cite{Tikhonov:2002um} to refine the geometries and optimize their performance. These optimization efforts are the reason that enabled the widespread adoption of the standard triple-GEM configuration in most of the experiments. The Standard GEM (SGEM) configuration consists of a pitch size of 140~\textmu m, an outer and inner hole diameter of 70~\textmu m and 50~\textmu m, respectively, a Kapton thickness of 50~\textmu m, and a metal layer thickness of 5~\textmu m \cite{Sauli:2016eeu}. GEM detectors with such configurations offer advantages such as low discharge probability, high gain, good spatial and temporal resolution, high rate capability, radiation hardness, design flexibility, and cost-effective manufacturing. Due to these characteristics, this standard configuration has proven highly effective in tracking and triggering applications in high-energy physics experiments. As a result, major experimental facilities such as the Large Hadron Collider (LHC)~\cite{Buonsante:2025lvd}, Common Muon and Proton Apparatus for Structure and Spectroscopy (COMPASS)~\cite{Ketzer:2004jk}, and Relativistic Heavy Ion Collider (RHIC)~\cite{Simon:2008qk, Azmoun:2018ail} have already implemented GEM detectors in triple-layered form. However, with upcoming high-luminosity collider experiments like High Luminosity Large Hadron Collider (HL-LHC) \cite{Hoepfner:2024vse} and Electron-Ion Collider (EIC) \cite{AbdulKhalek:2021gbh}, further advancements in GEM detector technology are essential. 

Further, to improve the performance of GEM detectors for high-luminosity experiments, one approach has recently been taken to improve the position resolution by reducing the pitch size of stacked GEM detectors. It has been reported that experimental investigations on a Fine Pitch GEM (FGEM) detector with a reduced pitch size of 90~\textmu m in a triple-GEM configuration show up to a 15\% improvement in spatial resolution \cite{Flothner:2024qev}. This improvement supports the hypothesis that reducing the GEM pitch enhances gain before saturation and discharge while improving spatial resolution. Encouraged by these findings, further reductions in pitch size were attempted to 60~\textmu m, leading to the development of an even finer GEM structure called Fine Thin Pitch GEM (FTGEM) \cite{EPDT_Seminar,cern2023seminar}. However, in contrast to assumptions, the FTGEM detector's spatial resolution did not show any improvement with pitch size reduction. Instead, the resolution was similar to that of the 90~$\mu$m pitch detector. It disputes the hypothesis coined. This discrepancy may have arisen due to the drift electric field, the induction electric field, and other geometrical parameters that are optimal for the standard GEM configuration, which might not be suitable for reduced pitch size GEM detectors. Therefore, although pitch size reduction has shown advantages, its full potential requires a more thorough examination of charge transport phenomena through simulation.

Since a single GEM detector is the building block of a multi-layered GEM configuration, understanding the behaviour of a single GEM under various circumstances becomes crucial. The electron multiplication occurs at multiple stages in a multi-layered GEM system and is highly influenced by the geometrical design of a single GEM. Additionally, pitch size and hole diameters directly impact key performance parameters. Therefore, analyzing the response of a single GEM with a smaller pitch size is essential before optimizing a multi-layer configuration. A deeper understanding of reduced-pitch GEM behaviour on transport properties will help refine detector designs, ensuring better performance and stability under extreme operating conditions.

\begin{figure*}[h!]
    \centering
    % First subfigure
    \begin{subfigure}[b]{0.3\textwidth} 
        \centering
        \includegraphics[width=\textwidth, height=5cm]{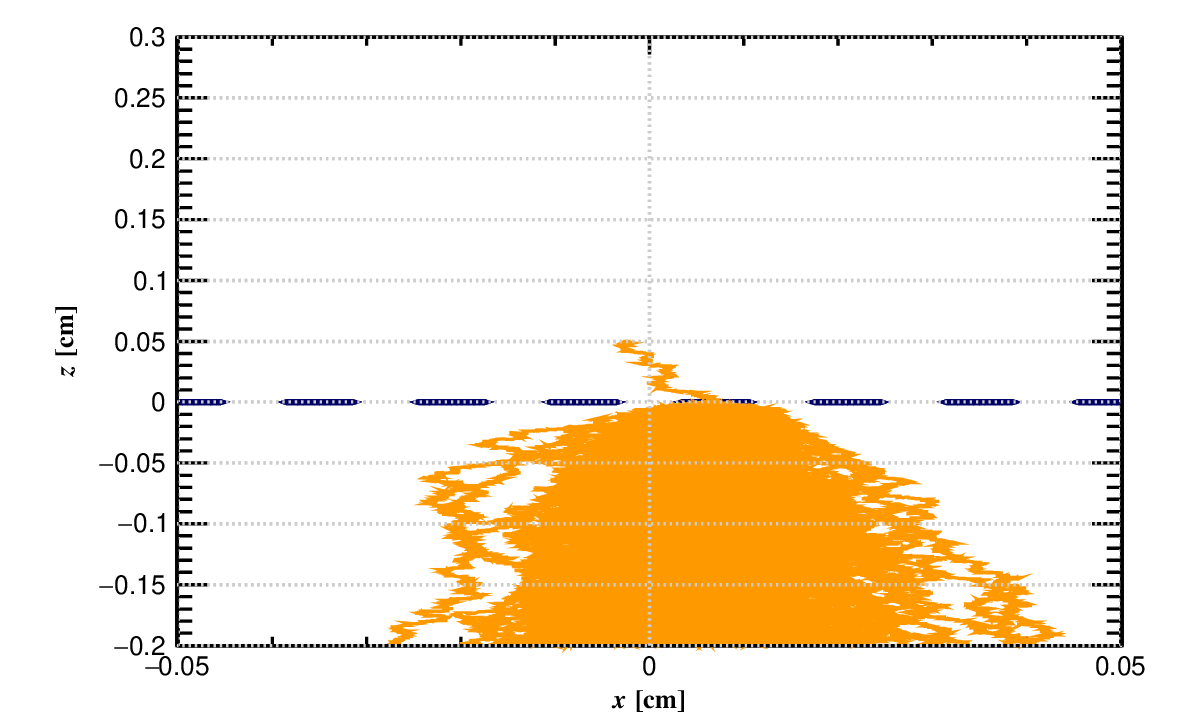}  
        \caption{}
        \label{fig:aval140}
    \end{subfigure}
    \hfill
    % Second subfigure
    \begin{subfigure}[b]{0.3\textwidth}  
        \centering
        \includegraphics[width=\textwidth, height=5cm]{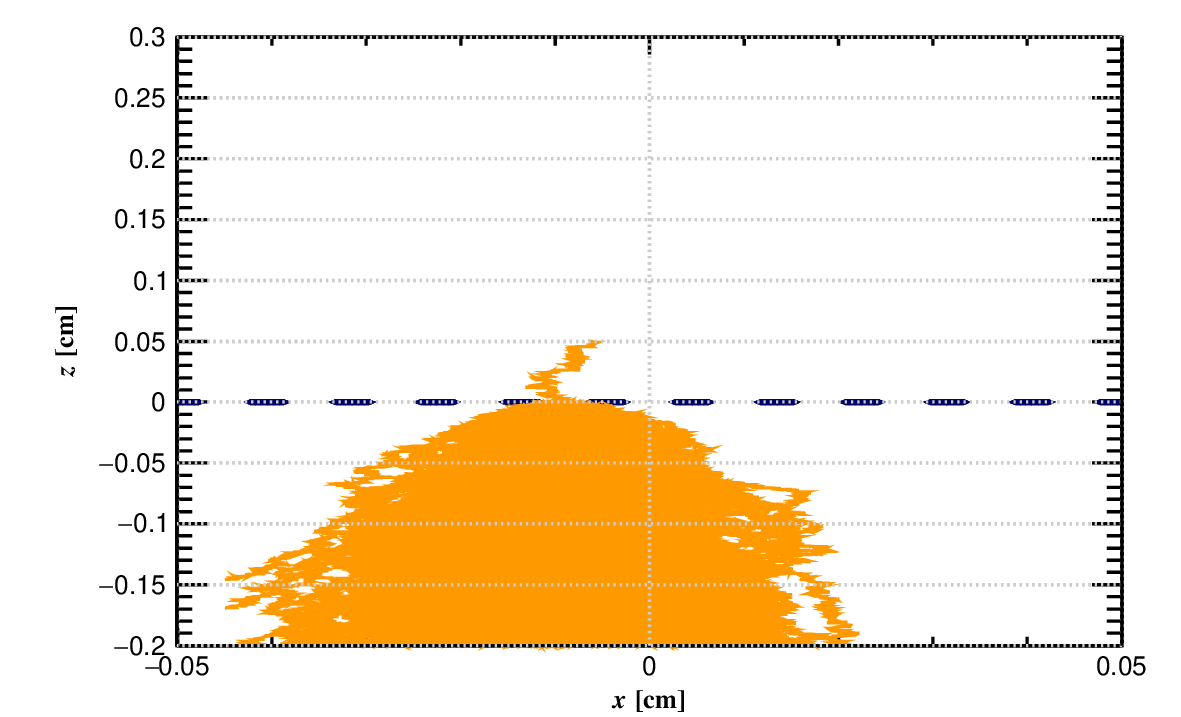}
        \caption{}
         \label{fig:aval90}
    \end{subfigure}
    \hfill
    % Third subfigure
    \begin{subfigure}[b]{0.3\textwidth}  
        \centering
        \includegraphics[width=\textwidth, height=5cm]{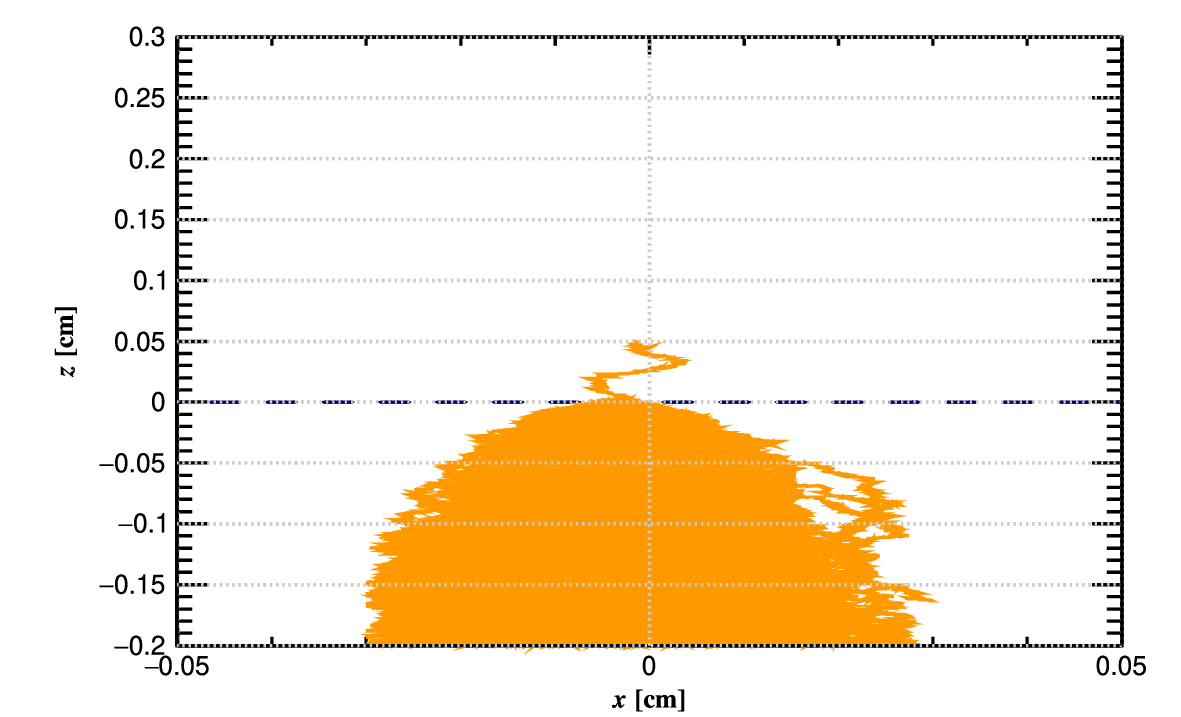}
        \caption{}
        \label{fig:aval60}
    \end{subfigure}

    \caption{Simulated drift lines along with 2D (XZ) distributions of a single electron in (a) SGEM, (b) FGEM, and (c) FTGEM.}
    \label{Fig1}
    
\end{figure*}

This article presents a detailed simulation analysis to understand the impact of pitch size reduction on detector performance. This work's initial focus is modelling a single standard GEM configuration to validate the simulation framework. To achieve this, we compare the simulated results with previously reported experimental measurements \cite{Bachmann:2000az}. Additionally, several other studies \cite{Sauli:2016eeu, Bachmann:1999xc, Jung:2021cvz, gem:2023amp, Wang:2015exp, Bencivenni:2002iw}, including both simulation-based and experimental works, have been reviewed to ensure consistency and reliability of the simulation approach.

After successfully validating the simulation framework, we examined the effects of reducing the pitch size to 90 µm and 60 µm on key performance parameters such as effective gain, position resolution, and electron transparency. These specific pitch sizes were motivated by experimental studies \cite{Flothner:2024qev, EPDT_Seminar, cern2023seminar}. We compared their performance relative to the standard configuration to assess the impact of pitch size reduction.

Among various performance parameters, the effective gain is particularly significant as it determines the total number of electrons collected at the induction plane from those generated in the drift region \cite{Wang:2022icv}. As effective gain is beneficial in quantifying the number of finally collected electrons, it is more worthy of consideration than intrinsic or real gain, representing the total electrons generated. The effective gain is always lower than the real gain due to factors such as electron attachment at the cathode, a positive ion cloud in the drift region and electron back-scattering. The reduction in pitch size is vital in determining these factors. Therefore, it is essential to understand the correlation between effective gain and pitch size.

Electron transparency is another performance parameter, which represents the fractions of the electrons that exit from the GEM holes electrode and finally get collected at the induction plate. Electron transparency can be expressed as the ratio of effective gain to real gain \cite{Wang:2022icv}.

The spread of electrons significantly affects position resolution, making it another crucial performance parameter \cite{cern2023quadruplefoils}. The increased spread of electrons at the induction plane complicates particle tracking. With the increase in collision rates, tracking spatially and temporally closed interactions will be much more difficult. Hence, the minimal spreading of electrons at the induction plane favours the track reconstruction.

Therefore, we have systematically analyzed reduced pitch size GEM detectors and compared their performance parameters with the standard GEM configurations by varying geometrical and electric field parameters.

This article is structured as follows: Section \ref{model} introduces the simulation framework and methodologies. Section \ref{simresults} presents simulation results from electron transport. This section is divided into seven sub-sections. Subsection \ref{ss_validity} discusses the validation of the simulation framework by comparing the simulated results with previously published experimental measurements~\cite{Sauli:2016eeu, Bachmann:1999xc}. Subsections \ref{ss_vgem} to \ref{ss_gas} analyze the variation of effective gain, electron transparency, and position resolution concerning varied pitch size single GEM detector with GEM potential $(\Delta \text{V}_\text{GEM}$), Drift Electric Field $(\text{E}_\text{D})$, Induction Electric Field $(\text{E}_\text{I})$, Drift Gap $(\text{X}_\text{D})$, Induction Gap $(\text{X}_\text{I})$, and gas composition. Finally, Section \ref{summary} concludes the paper and outlines future research directions. 

\section{Simulation Framework and Methodologies}\label{model}
%%\label{}\bibitem{SAULI20162}

Garfield++ \cite{Garfield++}, a potent tool for simulating semiconductors and gaseous detectors, has been used to carry out the simulation. While the original version of Garfield is in Fortran, Garfield++ is an enhanced version written in C++. Both versions are built within the ROOT framework \cite{ROOT} and have the same core functionality. 

Garfield++ includes numerous visualization classes for visualization processes like plotting drift lines, designing the layout of the detector, and making a contour plot of electrostatic potential. However, Garfield++ lacks classes for solving complex electric fields, which can be addressed by incorporating commercial software like ANSYS~\cite {ANSYS} and COMSOL~\cite{COMSOL}. We have utilized ANSYS as a finite element solver for our simulation study.
\begin{table}[ht]

%\hspace{0.7pt}
\centering
\begin{tabular}{|c|c|c|c|}
\hline
\textbf{Geometrical Parameters} & \textbf{SGEM } & \textbf{FGEM} & \textbf{FTGEM} \\ 
\hline
Pitch [$\mu$m] & 140 & 90 & 60 \\ 
Kapton Thickness [$\mu$m] & 50 & 50 & 25 \\ 
Metal layer [$\mu$m] & 5 & 5 & 5 \\ 
Inner diameter [$\mu$m] & 50 & 40 & 25 \\ 
Outer diameter [$\mu$m] & 70 & 55 & 30 \\ 
Rim diameter [$\mu$m] & 70 & 55 & 30 \\ 
\hline

\end{tabular}
\caption{GEM geometry and configuration with a drift gap of 3 mm, an induction gap of 2 mm, drift field of 2 kV/cm, induction field of 3.5 kV/cm, Ar: CO$_{2}$ (70:30) gas mixture, and a Penning transfer ratio (r$_\text{p}$) of 0.57 for Standard GEM (SGEM), Fine Pitch GEM (FGEM), and Fine Thin Pitch GEM (FTGEM).}\label{table}
\end{table}

The standard GEM (SGEM) is designed with a hexagonal pattern and features biconical holes. In addition to the metal contacts at the Kapton layer, both the drift and induction electrodes support these contacts. These metal contacts are essential for maintaining the necessary potential distribution for effective operation.

The fundamental geometric structure of the GEM system, created in ANSYS, consists of three key regions: the drift region, where primary ionization occurs; the avalanche region, where electron multiplication takes place; and the induction region, where multiplied electrons drift toward the readout plane. The 3D geometry constructed has a fine mesh that enables accurate computation of field maps. 

In our simulation, we have designed three distinct GEM detectors with varying pitch size and hole diameter, as detailed in Table \ref{table}. The unit cell is created in ANSYS and then imported into Garfield++. Garfield++ can mirror the unit cell to create a larger detector due to the symmetrical geometric design. A schematic representation of the GEM detector system, illustrating the different regions and dimensions, is shown in Figure \ref{Schematic}.

For each geometric configuration, ANSYS generates four key output files: ELIST (comprising a list of geometrical elements), NLIST (comprising a nodal information list), MPLIST (comprising a list of material properties), and PRNSOL (comprising a potential and response node list). Subsequently, these files are fed as input into Garfield++ to solve the dynamics of electron transport in the detector.

Garfield++ interfaces with three essential tools: Magboltz \cite{MAGBOLTZ}, Heed \cite{HEED}, and neBEM \cite{neBEM}. Magboltz is crucial for simulating electron transport properties in a gaseous mixture; Heed calculates the ionization pattern produced by the relativistic charge particle in the detector medium, and neBem aids in solving field maps. Notably, our simulation is based on use of the Magboltz class, which is pivotal in determining effective gain, electron transparency, and position resolution. A single electron's microscopic tracking is activated using the class AvalancheMicroscopic.

An electron with an energy of 0.5 eV is accelerated from 0.5 mm above the GEM detector, which initiates the simulation. The Monte-Carlo simulation has been performed based on the parameters outlined in Table~\ref{table}. The simulation is run in batch mode for 1000 events, as the results were consistent at these events. The higher event simulation was restricted due to high computational time. The penning transfer (r$_\text{p}$) to incorporate energy transfer for a 70:30 Ar: CO$_\text{2}$ mixture is 0.57. However, the simulation during the analysis focused on gas composition incorporates variation in the gaseous mixture, and accordingly, the magnitude of penning transfer has been considered based on the reported article~\cite{Sahin:2014haa}. The atmospheric pressure and temperature considered during the simulation is 1 atm and 293K, respectively.

\begin{figure}
	\centering 
	\includegraphics[width=1.0\linewidth]{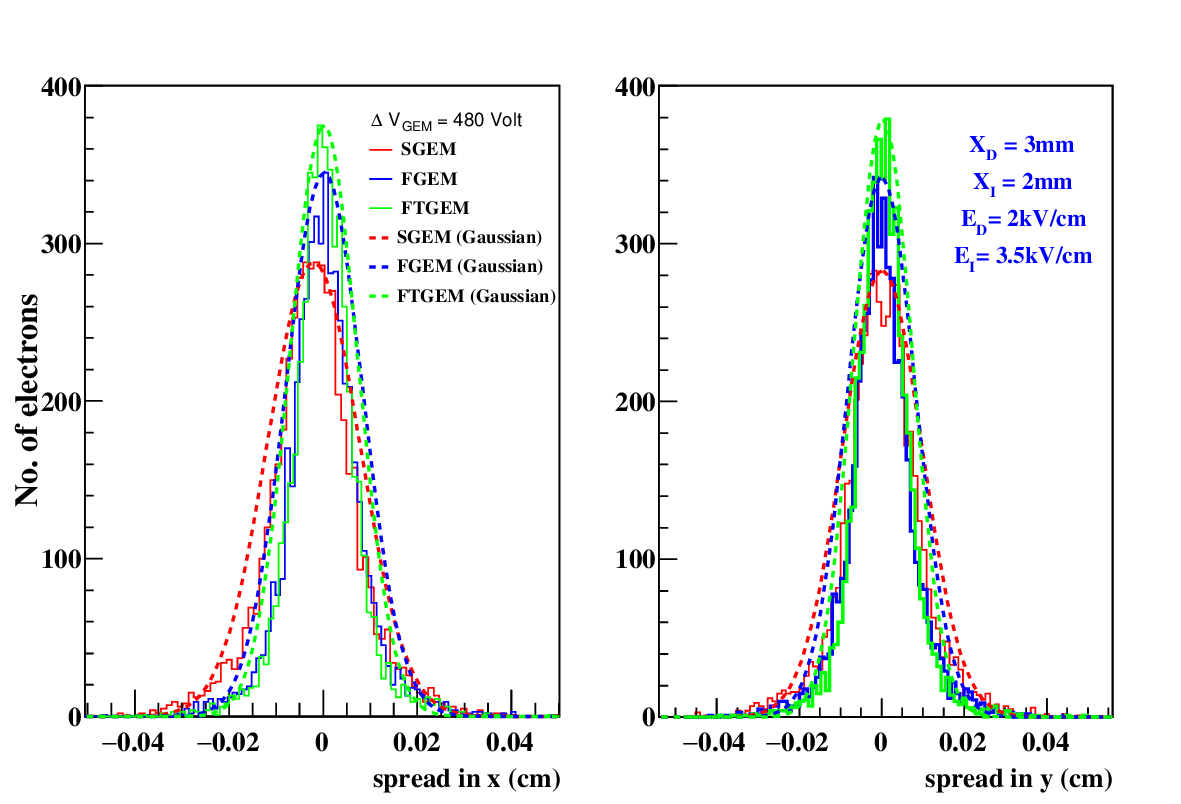}	
	\caption{Distribution of electrons in X coordinate and Y coordinate with a Gaussian fit.} 
	\label{fig_3}
\end{figure}

\begin{table*}[ht]

%\hspace{0.7pt}
\centering
\begin{tabular}{|c|c|c|c|c|c|c|}
\hline
\hline
\textbf{GEM Type} & \textbf{$\text{N}_{\text{Total}}$(e)} & \textbf{$\text{N}_{\text{Induction}}$(e)} & \textbf{$\text{N}_{\text{Kapton}}$(e)} & \textbf{$\text{N}_{\text{L.M.}}$(e)} & \textbf{$\text{N}_{\text{U.M.}}$(e)} & \textbf{$\text{N}_{\text{Others}}$(e)} \\ 
\hline
SGEM & 403 & 149 (36.97 $\%$) & 28 (6.95 $\%$) & 0 & 131 (32.51 $\%$) & 95 (23.57 $\%$)\\ 
FGEM & 698 & 187 (26.79 $\%$)& 75 (10.74 $\%$) & 0 & 264 (37.82 $\%$)& 172 (24.64 $\%$)\\ 
FTGEM & 6757 & 1301 (19.25 $\%$) & 603 (8.92 $\%$)& 0 & 3560 (52.68 $\%$) & 1293 (19.14 $\%$)\\ 

\hline
\hline

\end{tabular}
\caption{Record of the number of avalanche electrons in different regions with their percentage distribution for GEM detectors of different pitch sizes, simulated at 480 V. Here, $\text{N}_{\text{Total}}$ represent the total number of electrons generated; $\text{N}_{\text{Induction}}$ represents the total number of electrons collected at the induction plane; $\text{N}_{\text{Kapton}}$ represents a total number of electrons in Kapton layer; $\text{N}_{\text{U.M.}}$ represents the total number of electrons in the upper GEM electrode, $\text{N}_{\text{L.M.}}$ represents the total number of electrons in lower GEM electrode, and $\text{N}_{\text{Others}}$ represents the number of electrons that was collected elsewhere.}\label{table_record}
\end{table*}

\section{Simulation results from electron transport \label{simresults}}

 The drift lines of electrons, passing through the intense electric field region of the GEM detector with pitch sizes of 140 µm, 90 µm, and 60 µm, can be viewed in Figures \ref{fig:aval140}, \ref{fig:aval90} and \ref{fig:aval60}, respectively. The hole density rises as the pitch size decreases, producing denser drift lines that signify more avalanche events. This enhancement of avalanche events is due to increased electric fields at relatively narrower regions of holes. The higher density of outgoing drift lines from the holes favours higher avalanche multiplication corresponding to reduced pitch size. Furthermore, our simulation suggests a finer sampling of electron clouds from reduced-pitch GEM detectors as they approach the induction electrode due to smaller pitch sizes and narrower holes.
 
The position resolution is assessed based on the distribution of electrons on the induction plane \cite{cern2023quadruplefoils}. The X and Y spreads are defined based on the distribution of electrons in a spatial coordinate system. These spreads are determined by filling a histogram, which carries the information of the difference in reconstructed events corresponding to the final position at the induction electrode and the true events corresponding to the initial position of the avalanche electrons. 

As depicted in Figure \ref{fig_3}, the X and Y spread distributions of electrons on the induction plane show a Gaussian nature for all 3 GEM configurations. Additionally, we can observe that pitch size reduction leads to reduced dispersion of electrons on the induction plane. These Gaussian distributions of electrons are fitted using a peak-fitting Gaussian function as given by Equation \ref{eq:gauss} \cite{Lan-Lan:2013dxa}. 

\begin{equation} \label{eq:gauss}
y = \left( \frac{A}{w \sqrt{\pi / 2}} \right) \exp \left( -2 \frac{(x - x_c)^2}{w^2} \right)
\end{equation}

Here, in this Equation, \( y \) indicates the function value, \( A \) indicates the Gaussian function amplitude, and \( w [\mu m] \) indicates the width of the distribution. The independent variable is represented by the variable \( x [\mu m] \) and the centre position of the Gaussian peak, also called the mean position, is indicated by \( x_c [\mu m] \). These parameter values can be extracted from the Gaussian distribution of electron spread. From Figure \ref{fig_3}, we can observe that the electron distribution on the induction electrode can be effectively fitted with the given Gaussian function. The Gaussian nature of the electron distribution is due to the transverse diffusion of the electrons ejected from the multiplicative holes. This transverse diffusion of the ejected electrons occurs as the drift lines tend to converge to the central drift lines, resulting in a Gaussian profile. The resultant position resolution can be computed using the following formula, which considers the standard deviations ($\sigma_x$, $\sigma_y$) from the X and Y spread fits of electron distributions. 

\begin{equation} \label{eq:sigma}
\sigma = \sqrt{\sigma_x^2 + \sigma_y^2}
\end{equation}

Table \ref{table_record} illustrates the distribution of electrons following the avalanche mechanism in the examined GEM detectors. From Table \ref{table_record}, it is clear that, although reduced pitch size GEM detectors collect a significant number of electrons at the induction electrode for the specified value of $\Delta \text{V}_\text{{GEM}}$, a higher percentage of these electrons gets attached to the upper metal (directed towards induction region). The simulated data shows negligible electron attachment at the lower GEM electrode across all GEM configurations. Furthermore, it can be observed that the FGEM configuration traps a relatively higher number of electrons compared to the FTGEM and SGEM configurations in the hole region. Considering these observations, we will briefly discuss the simulation results.

\iffalse
    
\begin{table*}[ht]

%\hspace{0.7pt}
\centering
%\renewcommand{\arraystretch}{1} % Adjusts the vertical height
\begin{tabular}{|c|c|c|c|c|c|c|}
\hline
\hline
\textbf{GEM Type} & \textbf{A} & \textbf{w} &\textbf{x_c} & \textbf{$\sigma$}&   \textbf{x (X-spread)} &  \textbf{x (Y-spread)}  \\ 
\hline
SGEM & 7.026 & 194.7 & -20.51 & 97.35 & -20.63 & 1.2\\ 
FGEM & 7.026 & 162.52 & 1.72 & 81.26 & 1.72 & 0.02 \\ 
FTGEM & 5.23 & 144.86 & 0.98 & 72.43 & 0.98 & 2.88 \\ 
\hline
\hline
\end{tabular}

\caption{...}\label{table_record_eqn}
\end{table*}
\fi

 \begin{figure}
    \centering
    \includegraphics[width=1.0\linewidth]{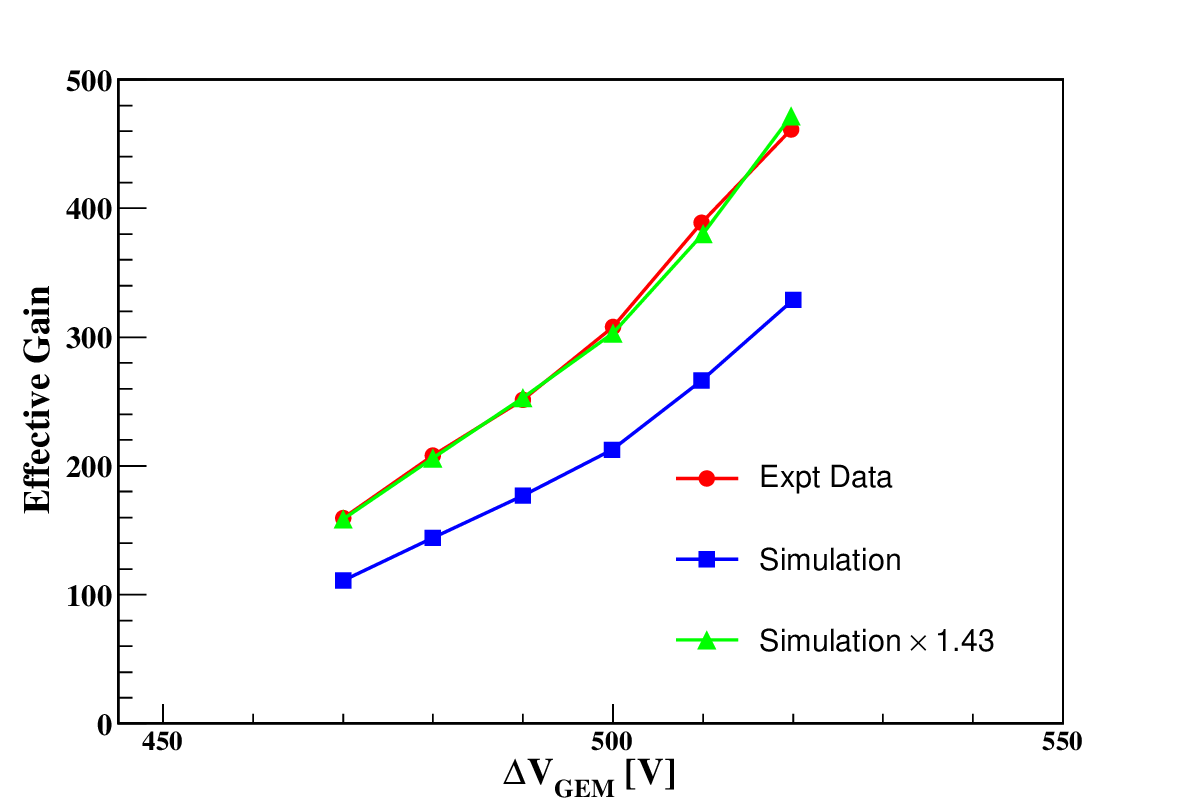}
        \caption{Variation of effective gain as a function of $\Delta V_{\text{GEM}}$ and comparison with experimental data~\cite{Bachmann:2000az}.}
    \label{fig_2}
\end{figure}

\subsection{Correlation Between Simulation and Experiment \label{ss_validity}}

We first validated the simulation framework against experimental measurements \cite{Sauli:2016eeu}, focusing on the variation in effective gain as a function of $\Delta V_{\text{GEM}}$. Several other results not included in this article have also been simulated and compared with experimental measurements \cite{Sauli:2016eeu}. These include the variation of real gain with $\Delta V_{\text{GEM}}$, the dependence of effective and real gain on hole diameter, and the variation of effective gain with $\Delta V_{\text{GEM}}$ for different gas compositions. The results revealed a similar trend to the experimental measurements \cite{Sauli:2016eeu}.

As shown in Figure \ref{fig_2}, the simulated effective gain closely follows the experimental trend as $\Delta V_{\text{GEM}}$ increases. It is estimated to be 1.43 times lower than the experimental effective gain, which aligns well with the results reported in simulation work \cite{Jung:2021cvz, gem:2023amp}. The discrepancies between simulation and experimental values may arise due to several factors. One possible reason is the absence of a photon feedback mechanism in Garfield++, as mentioned in Reference \cite{rahmani2018gem}. Additionally, other contributing factors include the imperfectness in designing an ideal GEM detector, ineffectiveness in calculating the electric field at mesh nodes using finite element software, charging up of the insulating foil \cite{Alfonsi:2012msp, Correia:2014vla}, the effect of Penning transfer ratio \cite{Azevedo:2016mer}, and various other external factors. Investigations are still ongoing to find out the exact causes of these differences.

\begin{figure}
        \centering
        \includegraphics[width=1\linewidth]{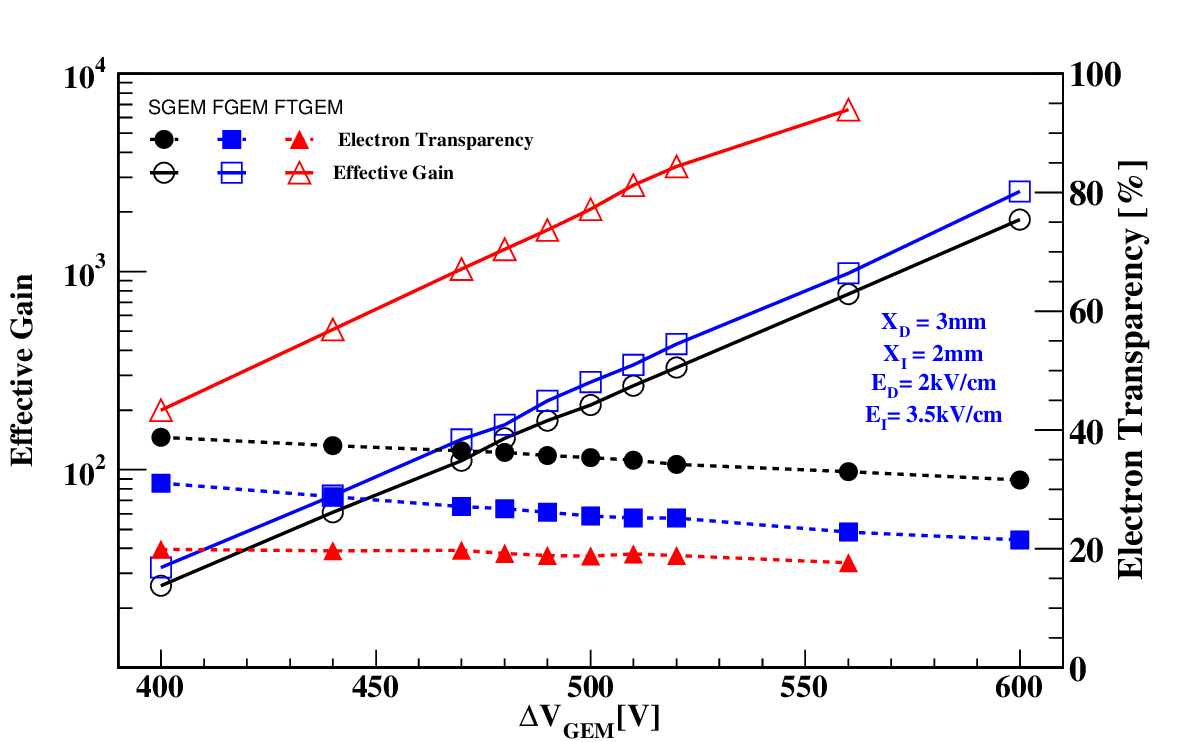}
        \caption{Variation of effective gain (open markers) and electron transparency (solid markers) as a function of $\Delta V_{\text{GEM}}$ for various pitch sizes.}
        \label{fig_4}
    \end{figure}

It has been observed from Figure \ref{fig_2} that incorporating a multiplication factor in the simulation improves agreement with experimental results. However, since the factors influencing this multiplication factor remain uncertain, and our objective is to provide a qualitative description of the results, we have not included this factor in further analysis.

\subsection{Effect of GEM potential ($\Delta V_{\text{GEM}}$) on effective gain, electron transparency, and position resolution. \label{ss_vgem}}

Figure \ref{fig_4} shows the variation of the effective gain (left Y axis) and electron transparency (right Y axis) for three different GEM configurations as a function of $\Delta V_{\text{GEM}}$. It shows that the effective gain of all three GEMs increases as the GEM potential increases. Further, the simulation results indicate that FGEM achieves nearly 1.25 times more effective gain than SGEM, while FTGEM provides almost 9 times more effective gain than SGEM with increasing $\Delta V_{\text{GEM}}$. The increase in effective gain with an increase in $\Delta V_{\text{GEM}}$ is attributed to the fact that avalanche rates increase due to the rise in electric field strength governed by $\Delta V_{\text{GEM}}$.

   \begin{figure}
       \centering
    \includegraphics[width=1\linewidth]{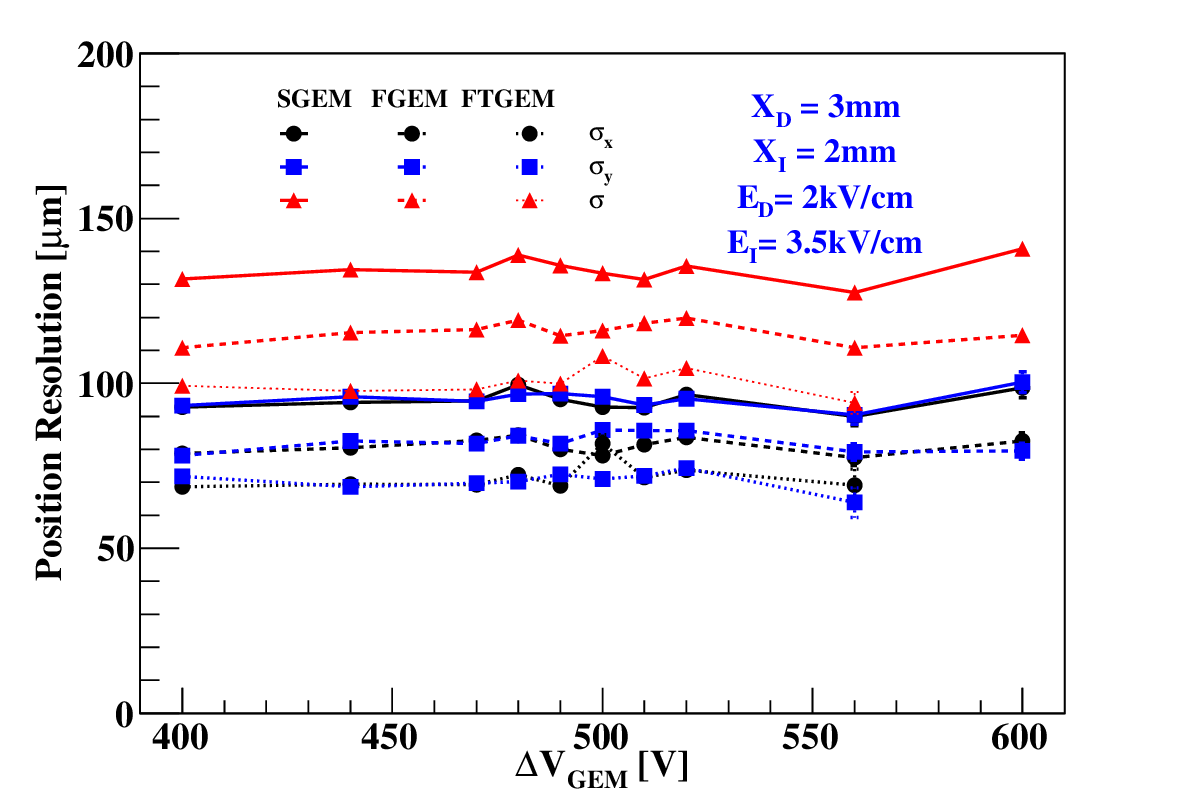}
     \caption{Position resolutions ($\sigma_{x}$, $\sigma_{y}$, and $\sigma$) as a function of $\Delta V_{GEM}$ for various pitch sizes.}
       \label{fig_5}
 \end{figure}
 
In contrast, from the same Figure \ref{fig_4}, we have observed that the electron transparency decreases with the increase in $\Delta V_{\text{GEM}}$. The reduction of electron transparency due to the rise in $\Delta V_{\text{GEM}}$ is likely due to enhanced electron attachments at the upper GEM metal. Furthermore, the reduced pitch size GEM detectors exhibits comparatively lower electron transparency. Despite the increase in effective gain with pitch reduction, the reduction in electron transparency is primarily due to the lower probability of electron transfer through the narrower holes. Factors such as ion back-flow, electron attachments, and accumulation of positive ions in the drift region significantly reduce electron transparency. The obtained electron transparency for different GEM configurations can be analyzed from electron transport properties illustrated in Table \ref{table_record}. For an SGEM, our simulated values of electron transparency are similar in magnitude to the reported article \cite{Jung:2021cvz}. 
 
The spread of electrons in the X and Y directions is quantified using $\sigma_{x}$ and $\sigma_{y}$. Figure \ref{fig_5} shows how \(\sigma_{x}\), \(\sigma_{y}\), and the overall position resolution \(\sigma\) vary with changes in \(\Delta V_{\text{GEM}}\). The resultant position resolutions \(\sigma\) remain nearly constant as \(\Delta V_{\text{GEM}}\) increases for the GEM configurations considered. Furthermore, the FTGEM configuration significantly enhances position resolution compared to both the FGEM and SGEM configurations. This improvement can be attributed to the narrower holes in the FTGEM design, which reduce the transverse diffusion of electrons. To better understand the position resolution profile, the data obtained from the different GEM configurations, as detailed in Table \ref{table_perform}, can provide valuable insights.

\begin{table}[ht]

%\hspace{0.7pt}
\centering
\begin{tabular}{|c|c|c|c|}
\hline
\hline
\textbf{GEM Type} & $\sigma_{x}$ [µm] & $\sigma_{y}$ [µm] & $\sigma$ [µm] \\ 
\hline
SGEM &  99.64$\pm$2.23 & 96.67$\pm$2.16 & 138.83$\pm$1.55  \\ 
FGEM &  84.28$\pm$1.88 & 84.14$\pm$1.88 & 119.09$\pm$1.33  \\ 
FTGEM & 72.32$\pm$1.62 & 70.24$\pm$1.57 & 100.81$\pm$1.13\\ 

\hline
\hline

\end{tabular}
\caption{Values of the parameters of position resolution at 480V.}\label{table_perform}
\end{table}

\subsection{Effect of Drift Electric Field ($\text{E}_\text{D}$) on effective gain, electron transparency, and position resolution}

 \begin{figure}
       \centering
    \includegraphics[width=1\linewidth]{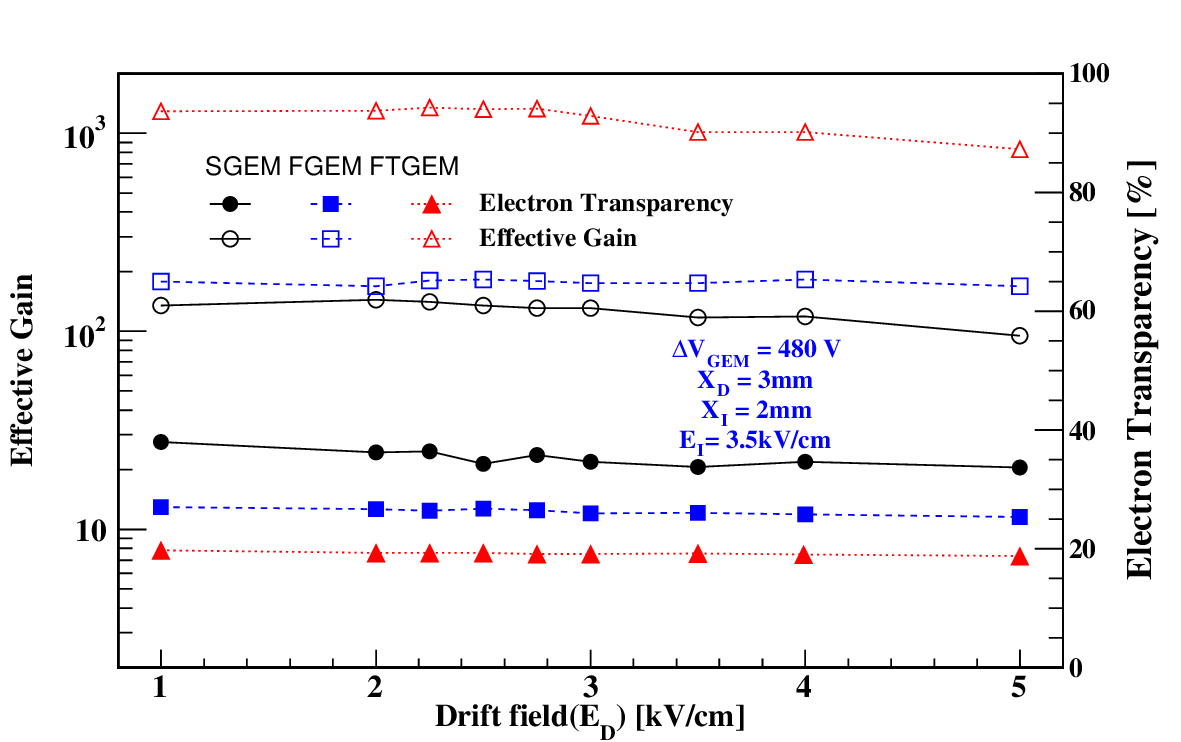}
     \caption{Variation of effective gain (open markers) and electron transparency (solid markers) as a function of $\text{E}_\text{D}$ for various pitch sizes at a constant $\Delta V_{GEM}$= 480 V.}
     \label{fig_4.2}
 \end{figure}
 
Figure \ref{fig_4.2} illustrates the variation of effective gain and electron transparency as a function of $\text{E}_\text{D}$  simulated at constant \(\Delta V_{\text{GEM}}\) of 480 V. It can be observed that with the increase in $\text{E}_\text{D}$ in SGEM, the effective gain increases to a maximum value up to 2.25-3 kV/cm and then decreases. In contrast, FGEM shows a slight deviation in effective gain, while the effective gain of FTGEM decreases above 2 kV/cm. The reduction of effective gain after reaching its peak value has also been observed for the case of thick GEM (THGEM) \cite{Hu:2021cel} and SGEM \cite{Bachmann:1999xc}. This reduction of effective gain after specific $\text{E}_\text{D}$ is possibly due to enhanced ion backflow, as discussed in the reported articles \cite{Tripathy:2021puz, SankarBhattacharya:2017pwk}.

In contrast, the electron transparency shows a decrease in trend with an increase in $\text{E}_\text{D}$. However, the rate of decline for FGEM and FTGEM is relatively lower than that for SGEM. This reduction is caused by enhanced electron loss due to attachment at the upper GEM electrode \cite{Sauli:2016eeu, Barbeau:2004zr}. Furthermore, SGEM demonstrates comparatively higher electron transparency than FGEM and FTGEM as \(\text{E}_\text{D}\) increases. As the electric field strength increases, the electrons gain more initial energy and are accelerated towards the holes. In GEM detectors with reduced pitch sizes, the electric field lines may become distorted at the edges of the holes. Consequently, ionized electrons may experience repulsion or deflections, further decreasing electron transparency.

   \begin{figure}
        \centering
        \includegraphics[width=1\linewidth]{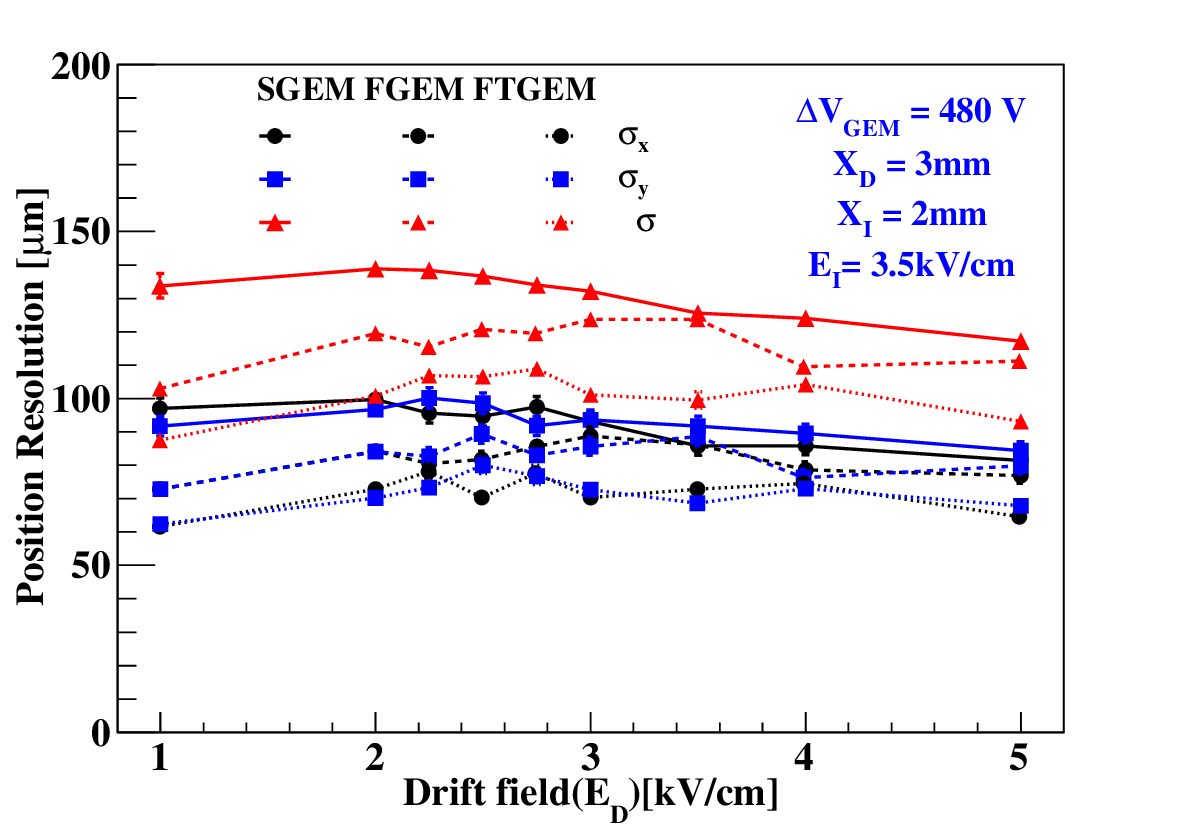}
        \caption{Variation of position resolutions ($\sigma_{x}$, $\sigma_{y}$ and $\sigma$) as the function of $E_{D}$ for various pitch sizes at a constant $\Delta V_{GEM}$= 480 V.}
        \label{fig_5.1}

    \end{figure}

Figure \ref{fig_5.1} shows the variation of position resolutions (\( \sigma_x \), \( \sigma_y \), and \( \sigma \)) as a function of $\text{E}_\text{D}$. We can observe that these values: \( \sigma_x \), \( \sigma_y \), and \( \sigma \)  are highest for SGEM, followed by FGEM, and lowest for FTGEM, indicating that FTGEM provides relatively good position resolution. Furthermore, the position resolution of GEM detectors improves as $\text{E}_\text{D}$ increases. This enhancement is likely due to the rise in drift velocity in the drift region. The increase in drift velocity leads to finer sampling of electron clouds through the GEM holes, leading to reduced signal spread at the induction plane.

\subsection{Effect of Induction Electric Field ($\text{E}_\text{I}$) on effective gain, electron transparency, and position resolution}

\begin{figure}
        \centering
        \includegraphics[width=1\linewidth]{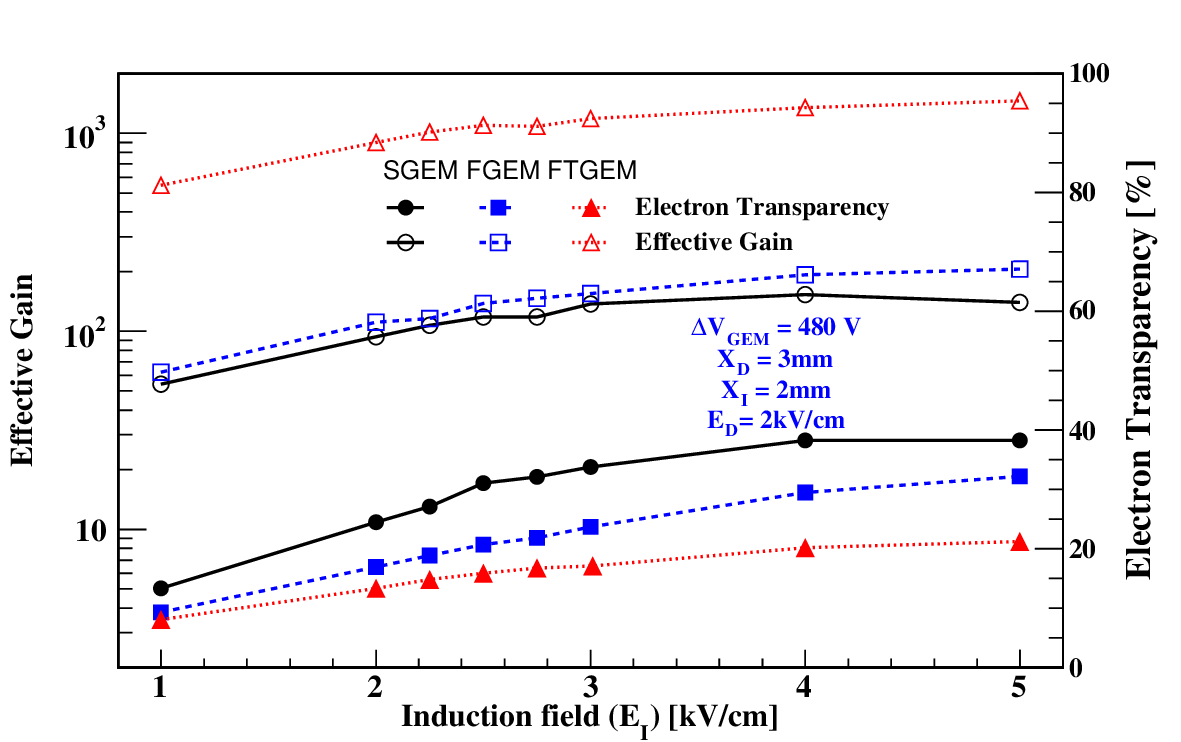}
        \caption{Variation of the effective gain (open markers) and transparency (solid markers) as a function of $\text{E}_\text{I}$ for various pitch sizes at a constant $\Delta V_{GEM}$= 480 V.}
        \label{fig_6}
    \end{figure}

Figure~\ref{fig_6} shows how the effective gain and electron transparency vary as a function of $\text{E}_\text{I}$. As $\text{E}_\text{I}$ increases, the effective gain rises for both FGEM and FTGEM, while SGEM experiences a slight decrease in effective gain after reaching 4 kV/cm. This increase in effective gain can likely be attributed to a stronger pulling force generated in the induction region, which hinders electron attachments to the GEM detectors and reduces electron recombination. This explanation is also consistent with the performance of GEM detectors with smaller pitch sizes. The SGEM has exhibited similar trends in previous experimental studies \cite{Bachmann:1999xc} as well as in reported simulations \cite{GEM_Garfield, Guedes:2003eq} when the value of $\text{E}_\text{I}$ is increased.

We can also observe that the electron transparency has enhanced with the increase in $\text{E}_\text{I}$. The SGEM, in this case, also holds better electron transparency than FGEM and FTGEM. At low values of $\text{E}_\text{I}$, electrons experience back scattering into the Kapton holes. Moreover, when $\text{E}_\text{I}$ is low, electrons trapped inside the holes may recombine with positive ions, decreasing electron transparency. However, at higher values of $\text{E}_\text{I}$, the likelihood of positive space charge formation in the drift region and the probability of electron trapping and backflow decrease. These combined effects contribute to the improved electron transparency by increasing $\text{E}_\text{I}$.

\begin{figure}
        \centering
        \includegraphics[width=1\linewidth]{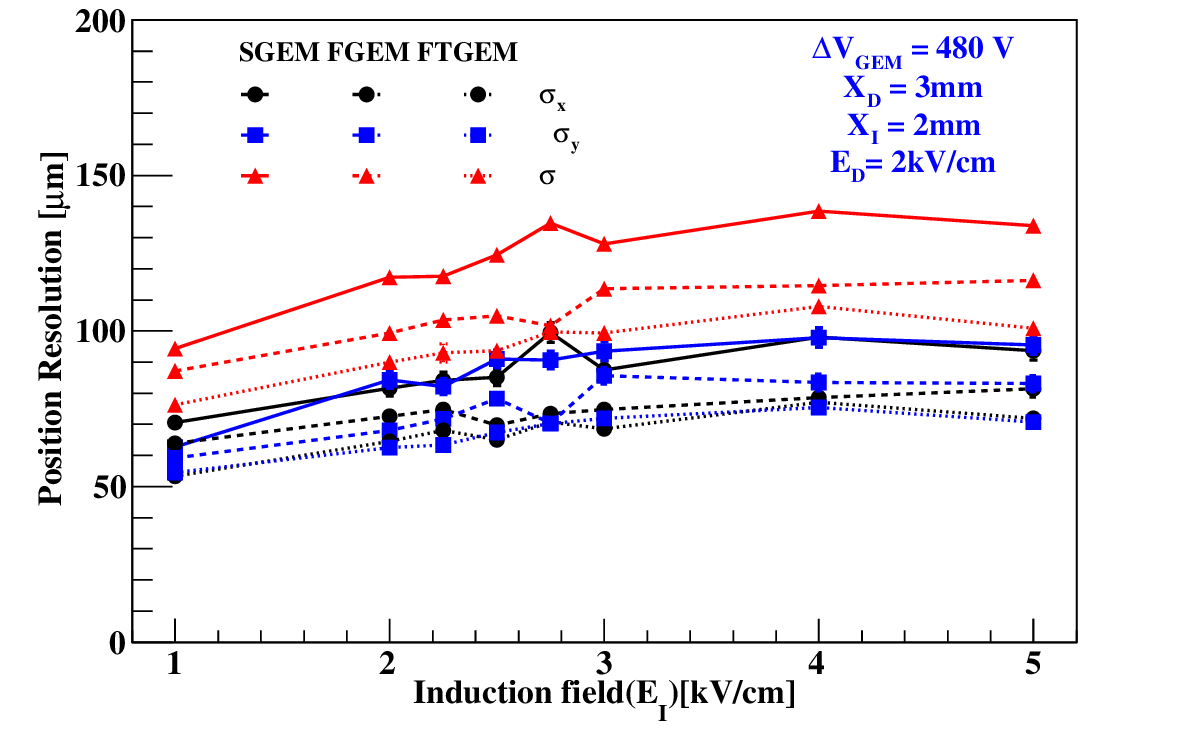}
        \caption{Variation of position resolutions ($\sigma_{x}$, $\sigma_{y}$ and $\sigma$) as a function of $E_{I}$ for various pitch sizes at a constant $\Delta V_{GEM}$= 480 V.}
        \label{fig_4.1}
    \end{figure}

Figure~\ref{fig_4.1} shows that the values of \( \sigma_x \), \( \sigma_y \), and \( \sigma \) increase as $\text{E}_\text{I}$ increases, suggesting a deterioration in position resolution. An increase in $\text{E}_\text{I}$ likely enhances the velocity of electrons from the GEM holes to the induction electrode, increasing the probability of angular deviations due to electron scattering. The increase in $\text{E}_\text{I}$ causes electrons to collide randomly with the composition mixture gases, increasing transversal spread and hence degradation of position resolution.

\begin{table} 
     \centering
     
\begin{tabular}{|c|c|c|c|}
\hline
Field Parameters & \textbf{SGEM } & \textbf{FGEM} & \textbf{FTGEM} \\ 
\hline
$\text{E}_\text{{D}}$ (kV/cm)  & 2.25 & 1 & 1 \\ 
$\text{E}_\text{{I}}$ (kV/cm) & 5 & 5 & 5 \\ 
%Effective Gain & 163 & 212 & 1382 \\
%E_{I}$ & 50 & 50 & 25 \\
%$E_{I}$ & 50 & 50 & 25 \\
%$E_{I}$ & 50 & 50 & 25 \\
\hline 
\end{tabular}

\caption{Optimized value of field parameters.}\label{table_2}
\end{table}

\subsection{Effect of Drift Gap ($X_\text{D}$) on effective gain, electron transparency, and position resolution}
\begin{figure}
        \centering
        \includegraphics[width=1\linewidth]{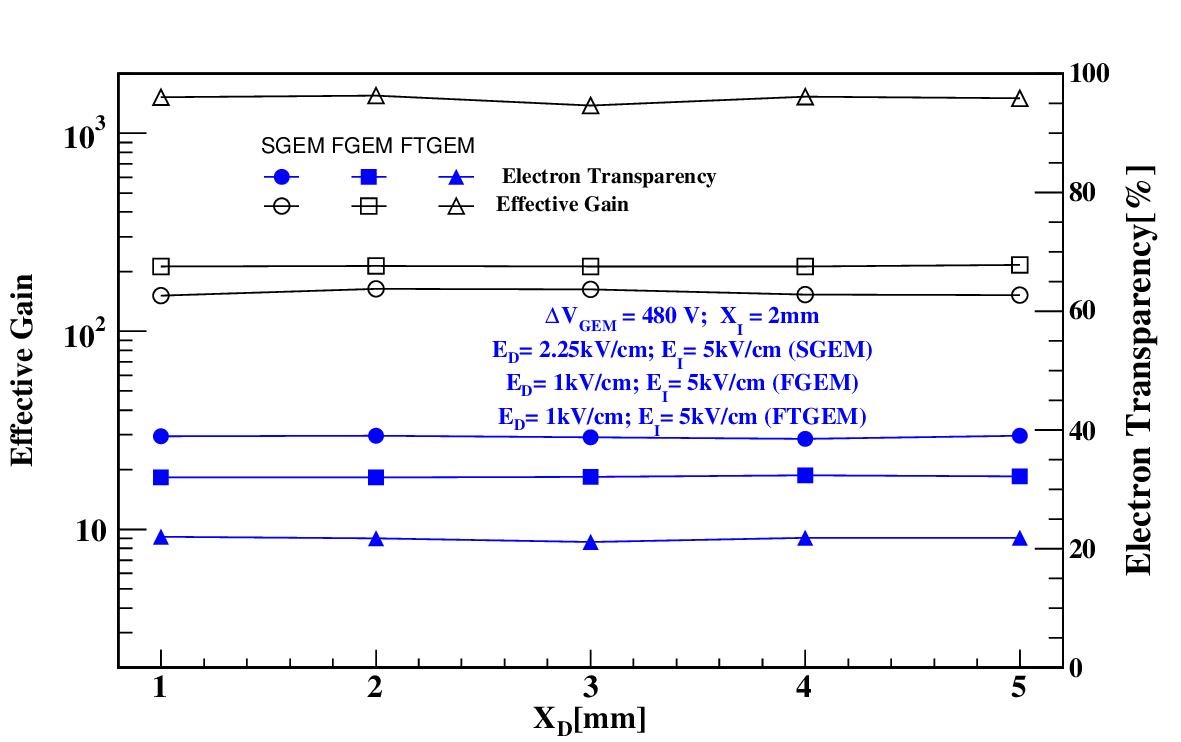}
        \caption{Variation of the effective gain (open markers) and electron transparency (solid markers) as a function of X$_\text{D}$ for various pitch sizes at a constant $\Delta V_{GEM}$= 480 V.}
        \label{fig_8}
    \end{figure}
    
    \begin{figure}
        \centering
        \includegraphics[width=1\linewidth]{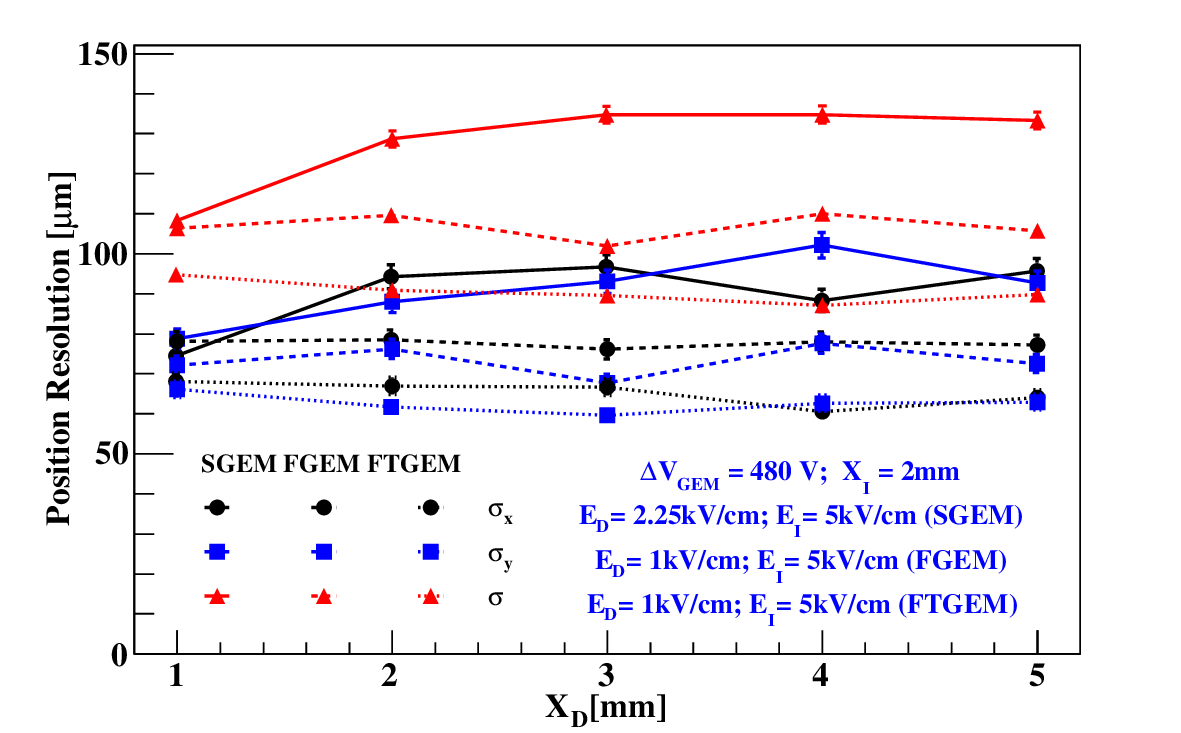}
        \caption{Variation of position resolutions ($\sigma_{x}$, $\sigma_{y}$ and $\sigma$) as a function of X$_\text{D}$ for various pitch sizes at a constant $\Delta V_{GEM}$= 480 V.}
        \label{fig_9}
    \end{figure}
After obtaining the optimized values of E$_\text{D}$ and E$_\text{I}$ for the specified GEM configurations, as shown in Table \ref{table_2}, simulations are conducted using these values across various pitch size GEM detectors to analyze the results related to changes in the width of X$_\text{D}$. This comparative analysis assesses the effects of increasing X$_\text{D}$ at a GEM potential of 480V. Figure \ref{fig_8} illustrates the relationship between effective gain and electron transparency as X$_\text{D}$ increases. Both electron transparency and effective gain are evident to remain relatively stable with increasing X$_\text{D}$. This stability occurs because changes in X$_\text{D}$ have a negligible effect on the avalanche mechanism, a critical process happening in the hole regions of GEM detectors. Although the drift gap may influence electron transport within the drift region, its impact on the variation of effective gain and electron transparency is minimal. 

In contrast, variations in the drift gap have a notable impact on the position resolution of GEM detectors. Figure \ref{fig_9} illustrates the variation of position resolutions with X$_\text{D}$. We can observe that the position resolution of SGEM decreases as the drift gap widens, while the position resolutions of FGEM and FTGEM slightly improve with an increasing drift gap. The decline in position resolution for SGEM is likely due to increased electron spread caused by transverse diffusion. Conversely, FGEM and FTGEM benefit from finer sampling of electron clouds, which is made possible by their smaller hole diameters. These factors lead to the observed improvement in position resolution as the drift gap increases.

\subsection{Effect of Induction Gap ($X_\text{I}$) on effective gain, electron transparency, and position resolution}
\begin{figure}
        \centering
        \includegraphics[width=1\linewidth]{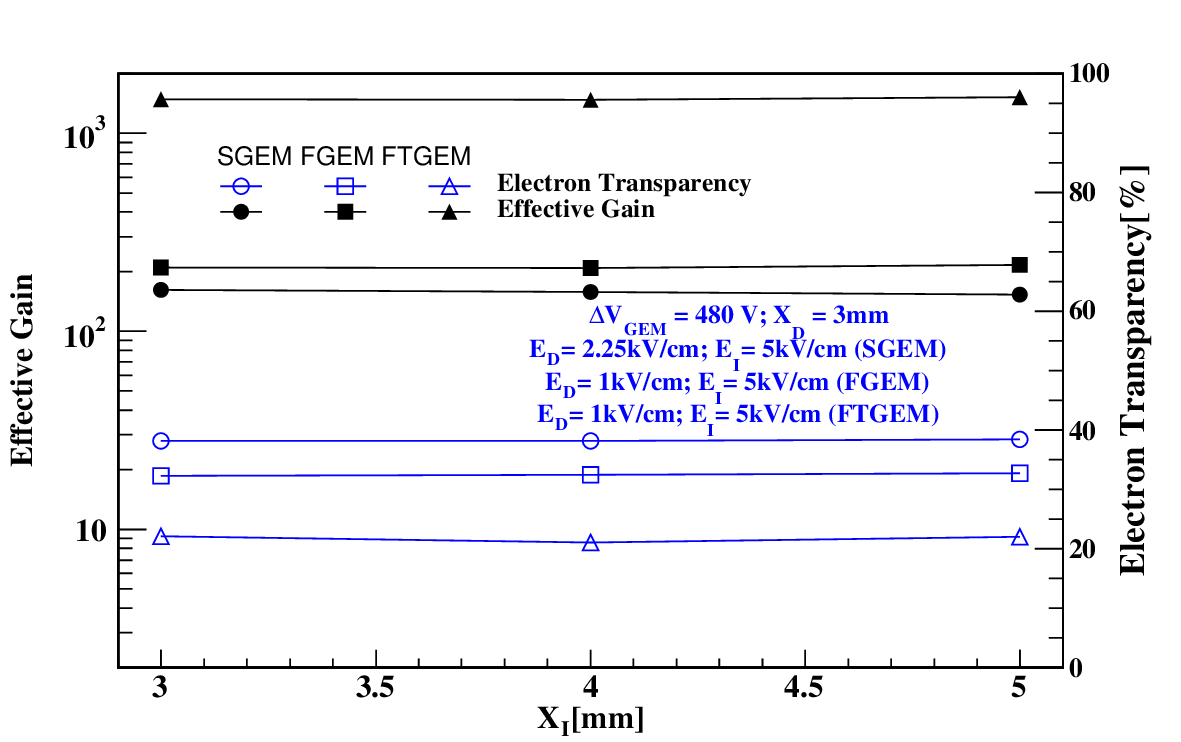}
         \caption{Variation of effective gain (solid markers) and electron transparency (open markers) as a function of X$_\text{I}$ for various pitch sizes at a constant $\Delta V_{GEM}$= 480 V.}
        \label{fig_8.1}
    \end{figure}
    \begin{figure}
        \centering
        \includegraphics[width=1\linewidth]{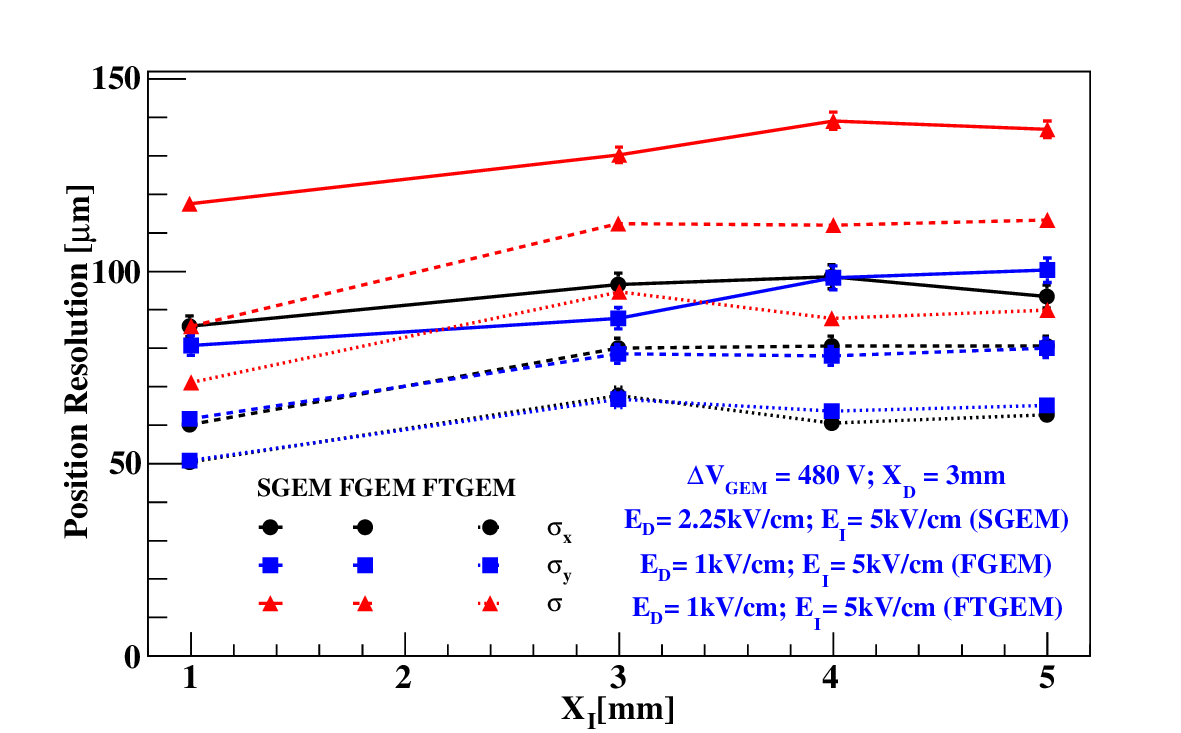}
        \caption{Variation of position resolutions ($\sigma_{x}$, $\sigma_{y}$, and $\sigma$) as a function of X$_\text{I}$ for various pitch sizes at a constant $\Delta V_{GEM}$= 480 V.}
        \label{fig_9.1}
    \end{figure}
The induction gap \( \text{X}_\text{{I}} \) for all three types of GEM detector configurations was varied from 3 mm to 5 mm, while the field parameter values were considered constant, as mentioned in Table \ref{table_2}. Figure \ref{fig_8.1} shows the variation of effective gain and electron transparency as a function of \( \text{X}_\text{{I}} \). We can observe that the width of the induction gap has a negligible effect on both the effective gain and electron transparency. As expected, the effective gain and electron transparency exhibit opposite trends with reductions in pitch size. The consistency of effective gain and electron transparency as the induction gap \( \text{X}_\text{{I}} \) increases indicates that there is no amplification mechanism occurring in the induction region.
    
Figure \ref{fig_9.1} illustrates how position resolution varies with an increase in \( \text{X}_\text{I} \). We observe differences in position resolution across various GEM configurations. Notably, there is a marked degradation in position resolution as the induction gap increases. This observation aligns well with findings from a previous study on a double GEM detector with a standard configuration \cite{Guedes:2003eq}. Our results support this conclusion, indicating that a larger induction gap negatively affects resolution. The primary cause of this degradation is the increased travel distance and time of avalanche electrons within a larger induction region, which leads to greater electron diffusion and, consequently, a reduction in position resolution.

\begin{table*}
     \centering
    
\begin{tabular}{|c||c|c|c|c|c|c|c|c|c|c|c|c|c|c|}
\hline
\hline 
\textbf{$CO_{2}$} (\%) & 1 & 2 & 3.5 & 5 & 7.5 & 10 & 15 & 20 & 25 & 30 & 35 & 40 & 45 & 50  \\ 
\hline
\textbf{$r_{p}$}  & 0.15 & 0.23 & 0.30 & 0.35 & 0.42 & 0.46 & 0.51 & 0.54 & 0.56 & 0.57 & 0.59 & 0.60 & 0.61 & 0.61  \\ 

%Effective Gain & 163 & 212 & 1382 \\
%E_{I}$ & 50 & 50 & 25 \\
%$E_{I}$ & 50 & 50 & 25 \\
%$E_{I}$ & 50 & 50 & 25 \\
\hline
\hline
\end{tabular}

\caption{The values of penning transfer ratio (r$_\text{p}$) for the given composition of CO$_{2}$ are obtained from the Reference \cite{Sahin:2014haa}.}\label{table3}
\end{table*}

\subsection{Effect of gas composition and penning effect on effective gain, electron transparency, and position resolution \label{ss_gas}}

\begin{figure}
        \centering
        \includegraphics[width=1\linewidth]{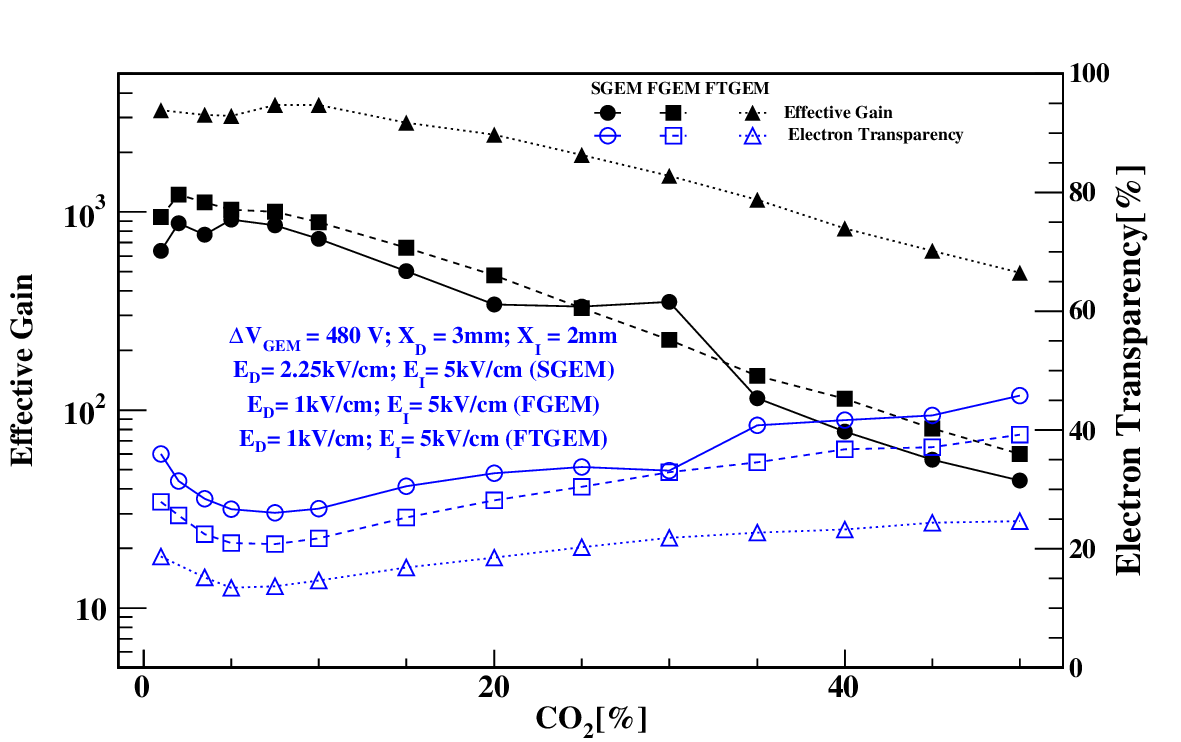}
        \caption{Variation of effective gain (solid markers) and transparency (open markers) as a function of CO$_{2}$ concentration for various pitch sizes at a constant $\Delta V_\text{GEM}$= 480 V.}
        \label{fig_8.2}
    \end{figure}

The effect of gas composition on effective gain and electron transparency has been analyzed for SGEM, FGEM, and FTGEM, as shown in Figure \ref{fig_8.2}, using Ar and CO$_{2}$ as the gas mixture. For an effective analysis, the corresponding values of penning transfer ratio (r$_\text{p}$) for each gas mixture have been taken from Reference \cite{Sahin:2014haa}. These values for corresponding CO$_{2}$ have been listed in Table \ref{table3}. As the CO$_{2}$ composition increases, the effective gain initially rises by up to 2-5\%, after which it starts to decline with further CO$_{2}$ addition. Notably, the difference in effective gain between FTGEM and SGEM becomes significantly larger at higher CO$_{2}$ concentrations. Furthermore, FGEM's effective gains remain close to the simulated values from SGEM as CO$_{2}$ concentration increases. One of the reasons for the decrease in effective gain after achieving its peak value with increasing CO$_{2}$ concentration is the increase in electron capture rate. Further, CO$_{2}$ is a quencher gas that captures excess free electrons, while Argon is a primary ionization source. Therefore, with an increase in CO$_{2}$ concentration, Argon concentration decreases to maintain the proportionality, favouring the reduction of effective gain. This trend was also reported earlier for SGEM \cite{Guida:2020ryb}. Operating at high voltages with CO$_{2}$ concentrations above 30\% can lead to higher gain. However, such operations are not recommended due to the increased risk of unintentional discharges
\cite{seidel2018microscopic}.

It is also worth noting that the effect of penning transfer can significantly alter the simulation results \cite{Sahin:2014haa}. This penning transfer describes the effect of an atom that gets excited to a higher energy state, by which it can effectively transfer part of its energy to the neighbouring quenching gas. This effect leads to additional electron creation. Therefore, we have considered this effect in our simulation analysis.

\begin{figure}
        \centering
        \includegraphics[width=1\linewidth]{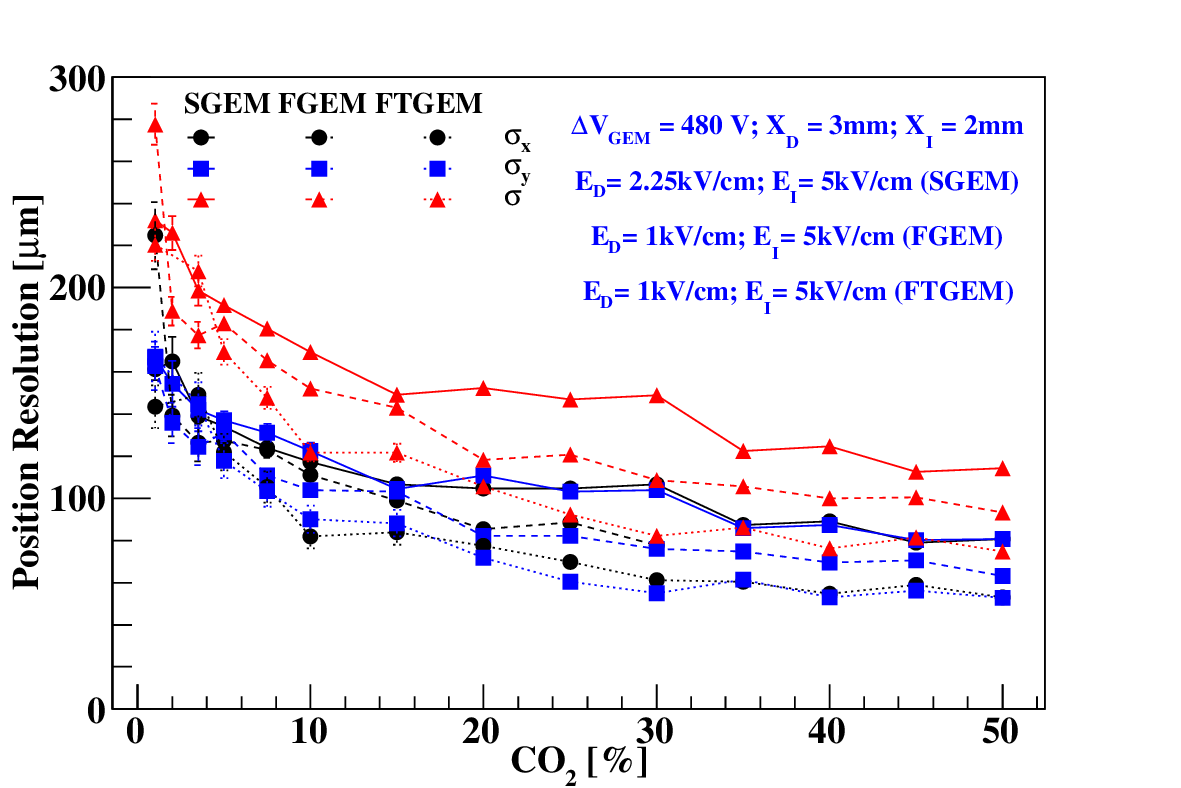}
        \caption{Variation of position resolutions ($\sigma_{x}$, $\sigma_{y}$, and $\sigma$) as a function of $CO_{2}$ concentration for various pitch sizes at a constant $\Delta V_\text{GEM}$= 480 V.}
        \label{fig_9.2}
    \end{figure}

Electron transparency, on the other hand, exhibits an opposite trend as it initially decreases by up to 5\%  $CO_{2}$ concentration and then increases beyond this point, which is visible in Figure \ref{fig_8.2}. The initial decrease in electron transparency for the considered GEM detectors is possibly due to enhanced electron attachments at the upper GEM metal. The higher concentration of CO$_{2}$ enhances the inelastic collision that reduces the drift velocity of electrons \cite{Swain:2025bsk}. This effect consequently leads to the reduction of electron transparency. On the other hand, the improvement of electron transparency after 5\% CO$_{2}$ composition for the given GEM configurations is possibly due to the increase in drift velocity caused by the suppression of electron diffusion. Lower transverse diffusion may result in focused electron transfer through the GEM holes, prohibiting electron scattering and attachments. Moreover, CO$_{2}$ is a non-polar molecule, which can contribute to providing a stable electric field as it does not show strong dipole interactions. These properties of $CO_{2}$ can lead to optimal electron transport through the holes, as a result of which electron transparency increases. 

The impact of gas composition on position resolution has also been investigated for SGEM, FGEM, and FTGEM. As shown in Figure \ref{fig_9.2}, position resolution improves with increasing CO$_{2}$ composition, with FTGEM exhibiting comparatively better spatial resolution than SGEM and FGEM at higher CO$_{2}$ percentages. A similar trend has been reported in previous studies for SGEM \cite{Guedes:2003eq}. This improvement may be attributed to the lower electron diffusion coefficient of CO$_{2}$ and better charge localization. However, it would not be appropriate to conclusively attribute the enhancement in position resolution solely to CO$_{2}$ composition, as several other factors, such as electron attachment losses, noise, and the readout circuit, which have not been included in this study, may also influence the results.

\section{Summary and Conclusions \label{summary}}

We have analyzed the comparative performance of GEM detectors for different pitch sizes: 140~$\mu$m (SGEM), 90~$\mu$m (FGEM), and 60~$\mu$m (FTGEM) in terms of effective gain, electron transparency, and position resolution. This study is motivated by experimental observations suggesting that a reduction in pitch size enhances both effective gain and position resolution in triple GEM configurations. 

%We first modelled a single GEM and examined the impact of varying pitch sizes to investigate the experimental trend. We further analyze these detectors by evaluating their performance under different geometrical and field configurations, including variations in $\Delta \text{V}_{\text{GEM}}$, $\text{E}_\text{D}$, $\text{E}_\text{I}$, $\text{X}_\text{D}$, $\text{X}_\text{I}$, and gas composition.  

The simulation results for GEM detectors with different pitch sizes indicate that FGEM and FTGEM, which have smaller pitch sizes and hole diameters, consistently achieve higher effective gain and better position resolution than SGEM detectors. However, although enhancement of effective gain and position resolution is favoured, their electron transparency remains comparatively low. Additionally, the simulation results for the considered GEM detectors highlight that decrease in $\text{E}_\text{D}$ and increase in $\text{E}_\text{I}$ favours greater effective gain and electron transparency, while enhancement in their position resolution at high values of $\text{E}_\text{D}$ and low values of $\text{E}_\text{I}$. Moreover, we investigated the impact of varying $\text{X}_\text{D}$ and $\text{X}_\text{I}$. We found that they have a negligible effect on effective gain and electron transparency while significantly contributing to better position resolution. Finally, we simulated the impact of gas mixtures on performance parameters. The simulation results show that increasing the CO$_2$ concentration enhances electron transparency and position resolution, while decreasing the CO$_2$ concentration leads to higher effective gain.

These findings suggest that small pitch size GEM detectors could be suitable for experiments requiring high gain and better position resolution. These findings also support the future uses of reduced pitch size with smaller hole diameters in multi-layered configurations. However, the abundant electron attachments at the upper GEM metal decrease the charge collection efficiency, requiring thorough examination to maximize the benefits of reduced pitch size GEM detectors. Moreover, simulation alone will not be sufficient to conclude performance parameters. Several additional factors, including flaws in the design of the simulated GEM geometry, finite element simulation limitations, electron attachment coefficients, discharge phenomena, and the impact of readout electronics, could cause disparities between simulated and experimental results. Future experiments focusing on variations of simulated parameters will be necessary to validate and reinforce these simulation outcomes.  

Despite the limitations, this work is still effective in providing qualitative descriptions of the influence of pitch size reduction on detector performance due to variations in various geometrical, field, and gas composition parameters. Considering this simulation result, we plan to evaluate their performance by exploring multi-layer GEM configurations stacked with different pitch-size GEM detectors. Understanding how different electrical and geometrical designs impact overall detector performance will be crucial for optimizing GEM technology. Such advancements will be significant in developing efficient GEM detectors for future collider experiments requiring high-precision measurements and high gain.   

\section*{Acknowledgements}
AK sincerely acknowledges the financial support from the Institute of Eminence (IoE), BHU grant - 6031. RG and SS acknowledge the financial support obtained from UGC under the research fellowship scheme in central universities.

%\tableofcontents



%%\end{thebibliography}

\end{document}